\documentclass[12pt]{article}

\usepackage{scicite}
\usepackage{mcite}
\usepackage{times}

\newenvironment{sciabstract}{%
\begin{quote} \bf}
{\end{quote}}

\usepackage{mathtools}
\usepackage{soul,xcolor}
\usepackage[makeroom]{cancel}
\sethlcolor{yellow}

\title{From Counterportation to Local Wormholes}

\author
{Hatim Salih$^{1\ast}$ \\
\\
\normalsize{$^{1}$Quantum Technology Enterprise Centre, HH Wills Physics Laboratory,} \\ \normalsize{University of Bristol, Tyndall Avenue, Bristol, BS8 1TL, UK}\\
\\
\normalsize{$^\ast$E-mail:  salih.hatim@gmail.com}
}

\date{}

\begin{document} 


\baselineskip24pt


\maketitle


\begin{sciabstract}
We propose an experimental realisation of the protocol for the counterfactual disembodied transport of an unknown qubit---or what we call counterportation---where sender and receiver, remarkably, exchange no particles. We employ cavity quantum electrodynamics, estimating resources for beating the classical fidelity limit---except, unlike teleportation, no preshared entanglement nor classical communication are required. Our approach is multiple orders of magnitude more efficient in terms of physical resources than previously proposed implementation, paving the way for a demonstration using existing imperfect devices. Surprisingly, while such communication is intuitively explained in terms of ``interaction-free'' measurement and the Zeno effect, we show that neither is necessary, with far-reaching implications in support of an underlying physical reality. We go on to characterise an explanatory framework for counterportation starting from constructor theory: local wormholes. Conversely, a counterportation experiment demonstrating the traversability of space, by means of what is essentially a 2-qubit exchange-free quantum computer, can point to the existence in the lab of such traversable wormholes.
\end{sciabstract}

\section*{Introduction}
``It is wonderful that we have met with a paradox. Now we have some hope of making progress,'' Neils Bohr once said. Paradoxes are puzzles that highlight some of the stranger aspects of physical theories, pointing to gaps in our present understanding. Their resolution, as Bohr enthused, often marks genuine progress. But never mind the ramifications of a future resolution of Schrodinger's cat paradox for instance, one just has to think of the advances this imagined feline physics companion has already instigated.

While a new physics paradox that we present here, about where a photon has or has not been within nested interferometers, throws up interesting considerations about the arrow of time, its resolution has a direct bearing on the possibility of counterfactual communication, i.e. sending information without exchanging particles \cite{Salih2013,*2013QuantumCommunication}---for long thought impossible. But if particles, as we establish using three different approaches, did not carry information across in such a scenario, then what did? And what does this say about the reality of the quantum state?---that mathematical construct that divides scientists as to whether it merely represents a state of knowledge or objective physical reality.

In an optical setting, we make the assumption that communication is explainable by one or more of the following: 1) detectable photons traversing the channel between the two communicating parties; 2) measurements carried out in between an initial quantum state and a final quantum state; and 3) an underlying physical state objectively existing prior to measurement. By categorically ruling out the first two in our scheme, we provide evidence that an underlying physical state, which we know from recent no-go theorems is uniquely represented by the quantum state, is what has carried quantum information across space.

Starting from the recently formulated constructor theory of information \cite{Marletto2014}, which shifts the focus from dynamical laws to questions about which tasks are possible and which are not and why, we go on to propose a general yet precise explanatory framework for exchange-free communication---motivated by an ambitious version of Maldacena and Susskind's ER=EPR conjecture \cite{Susskind,PhysRev.47.777,PhysRev.48.73}, where entanglement can connect not only regions of exotic blackhole configurations, but more generally ordinary regions of space.

Given the recent experimental demonstration of sending classical information without exchanging particles \cite{Cao2017,Salih2013,*2013QuantumCommunication}, we show that it is experimentally feasible to demonstrate sending not only classical but also quantum information this way, based on the generalised protocol, Salih14, for counterfactually transporting an unknown qubit \cite{Salih2016,*Salih2014b}. The protocol was the first to achieve the end goal of teleportation \cite{PhysRevLett.70.1895} without the basic requirements of previously shared entanglement and classical communication. At its heart is a distributed-across-space, counterfactual CNOT gate---the very first exchange-free gate \cite{Salih2016,*Salih2014b}. Since counterfactuality in this context means that the two spatially separated parties (with the control qubit at one party and the target qubit at the other) exchange no particles, we use the terms ``counterfactual'' and ``exchange-free'' interchangeably. We show how to implement this exchange-free CNOT based on cavity quantum electrodynamics \cite{Reiserer2015Cavity-basedPhotons}, and ideas inspired by electromagnetically induced transparency \cite{Mucke2010}. Our proposed realisation is remarkably tolerant to device and channel imperfections, and as the numerical simulations show, is implementable with current technology. This surprising two-qubit gate, combined with single-qubit operations, allows universal exchange-free quantum computation---across space that is---including the counterfactual transport of an unknown quantum state from one or more senders to one or more receivers: Counterportation.

\section*{The Paradox}
Let's get straight to the heart of the uncovered paradox, which we will first discuss generally in terms of the setup shown in Fig. \ref{fig:Paradox}, before framing the discussion in the specific context of counterfactual communication, revealing some of its more dramatic consequences. Consider the two outer interferometers, nested within each are two inner interferometers, shown in Fig. \ref{fig:Paradox}. All beam-splitters are polarising beam-splitters, transmitting horizontally polarised photons and reflecting vertically polarised photons, and all half-wave plates HWP's rotate polarisation by 45 degrees. The evolution of a photon between times $t_0$ and $t_4$, the first cycle, is identical to its evolution between times $t'_0$ and $t'_4$, the second cycle. Moreover, the photon is in the same state at times $t_0$, $t_4$, $t'_0$, and $t'_4$, in terms of both path and polarisation, provided that it is not lost to either of the detectors $D_A$. (As we will see shortly, this definition of a cycle is not arbitrary.) We want to know whether a photon detected at detector $D_0$ at the bottom was in any of the arms labeled C leading to the mirrors $MR_B$ on the right-hand side.

We follow the photon's evolution starting with the photon in arm S at the top at time $t_0$, H-polarised. The combination of the first half-wave plate HWP1 and plolarising beam-splitter puts the photon in an equal superposition of travelling along arm A, H-polarised, and along arm D, V-polarised. The combined action on the V-polarisd part by two successive half-wave plates HWP2's, one in front of each inner interferometer, is to rotate V polarisation all the way to H. This part of the superposition proceeds towards the first detector $D_A$. If $D_A$ does not click, then we know that the photon is instead in arm S at time $t_4$, H-polarised, having travelled along arm A. Exactly the same happens between times $t'_0$ and $t'_4$, which means that given it is not lost to $D_A$, the photon having taken the left-hand-side route will be in arm F at $t_{final}$, H-polarised, causing detector $D_0$ at the bottom to click.

An interesting approach for investigating whether a photon detected at $D_0$ at the bottom has travelled along any of the arms C, which common sense tells us should not be the case, is weak measurement on pre-selected and post-selected particles \cite{Aharonov1988}---an approach whose application has at times proven controversial in this context\cite{Danan2013,Salih2015,Danan2015,Vaidman2014a,Salih2014a,Englert2017,Peleg2018,*Englert2019}. The idea is to perform sufficiently weak measurements such that their effect on individual photons lies within the uncertainty associated with the observable measured, in this case the photon's path. When averaged over a sufficiently large number of photons, however, these measurements acquire definite, predictable values. One could, for instance, cause various mirrors in the setup to vibrate at different frequencies, as in reference \cite{Danan2013}, then check which of these frequencies show up in the power spectrum of classical light detected by a quad-cell photodetector $D_0$.

An intuitive way of predicting the outcomes of such weak measurements---at least to a first order approximation---is the two-state vector formulation, TSVF \cite{Aharonov1964,Aharonov1990,Danan2013}. According to the TSVF, each photon detected at detector $D_0$ is described by a forward-evolving quantum state created at the photon source at the top, and a backward-evolving quantum state created at detector $D_0$ at the bottom. Unless these two states overlap at a given point, in which case a well defined quantity related to the weak measurement, called the weak value, is nonzero, any weak measurement performed there will be vanishingly small. This is because any weak enough coupling to a variable $O$ of a pre- and post-selected system, such as the projection operator on the location of the mirror, results in an effective coupling to the weak value of the variable, $O_w \equiv \frac{\left\langle \Phi \right| O \left| \Psi \right\rangle}{\left\langle \Phi | \Psi \right\rangle}$, where $\left| \Psi \right\rangle$ is the forward-evolving pre-selected state and $\left\langle \Phi \right|$ is the backward-evolving post-selected state, using Dirac notation.

We can thus consider a weak measurement in arm C at time $t_2$, carried out on a beam of light by vibrating the corresponding mirror $MR_B$ at some frequency, which is then checked for in the power spectrum of light detected by quad-cell photodetector $D_0$. Starting from the photon source at the top, and following the photon's unitary evolution, the forward-evolving state is present in arm C. And starting from detector $D_0$ at the bottom, the backward-evolving state is also present in arm C. The weak value is nonzero and consequently a weak measurement in arm C at time $t_2$ is also nonzero.

Similarly, we can consider a weak measurement in arm C at time $t'_2$. Starting from the photon source at the top, the forward-evolving state is present in arm C. However, starting from detector $D_0$ at the bottom, the backward-evolving state is {\it not} present in arm C. The weak value is zero and therefore a weak measurement in arm C at time $t'_2$ is vanishingly small. Yet, in the absence of a weak measurement, the first outer cycle and the second outer cycle are identical as far as standard quantum mechanics is concerned---the photon undergoes the same transformations in each cycle, starting and finishing each in the same state.

The issue here is that weak measurement disturbs interference in the inner interferometers, beyond that due to device imperfections and the effect of the environment, both of which can be made negligible as in the experiment in \cite{Danan2013}, leading to a flux of V-polarised photons in arm D between times $t_3$ and $t_4$ ($t'_3$ and $t'_4$). Importantly, this flux finds its way to detector $D_0$ at the bottom only in the case of a weak measurement in the first outer cycle, as an artifact caused by the interference action of HWP1 {\it after} the end of the first outer cycle. In other words, the result of a weak measurement during the first outer cycle is not only dependent on what happens during that cycle, but also what happens {\it afterwards}. While it is remarkable that the TSVF can usually predict the outcome of such weak measurement experiments, it nevertheless manifests the inherent unusualness of the time-symmetric formulation of physics, where the present is not only dependent on the past but equally on the future\cite{Aharonov2011}. The formulation is too enticing to brush aside, yet its throwing away of the arrow of time comes at a cost \cite{Pusey2017}.

In a world where the future doesn't shape the past, one runs into trouble if nonzero weak-values at arms C are taken to mean the photon was necessarily there, a position advocated by Vaidman's weak-trace criterion \cite{Vaidman2013}. In the absence of a weak-measurement experiment, there is no photon leakage from the inner interferometers through arms D, except for that caused by the negligibly limited but otherwise unpredictable device- and environment-related imperfections acting on each cycle. Consequently, the evolution of the photon during the first cycle is approximately equivalent to its evolution during the second cycle, in which case saying the photon was in arm C for the first cycle but not for the second is unsettlingly paradoxical.

Our resolution of the paradox, from within the weak measurement framework, is based on the observation that the strong measurement by detector $D_A$ at the end of each outer cycle projects the state of the photon onto arm S, where we know it should be H-polarised. This is the post-selected state. Therefore for the first outer cycle, starting with the pre-selected state, where the photon is in S at time $t_0$ H-polarised, the forward-evolving state is present in arm C at time $t_2$ (and $t_3$). Starting with the post-selected state, however, where the photon is in S at time $t_4$ H-polarised, the backward-evolving state is {\it not} present in arm C at time $t_2$ (and $t_3$). The weak value is thus zero and a weak measurement will not find the photon there. Exactly the same applies for the second outer cycle, starting with the pre-selected state, the photon in S at time $t'_0$ H-polarised, the forward-evolving state is present in arm C at time $t'_2$ (and $t'_3$). Starting with the post-selected state, the photon in S at time $t'_4$ H-polarised, the backward-evolving state is {\it not} present in arm C at time $t'_2$ (and $t'_3$). The weak value is thus zero and a weak measurement will {\it not} find the photon there. Due to the disturbance caused by a weak measurement in C, and imperfections, the post-selected state in S at time $t_4$ (and $t'_4$) would have a small V-polorised component. Nonetheless, because the pre-selected forward evolving state and a post-selected V-polarised state in each cycle are ideally orthogonal, only a negligible amount of V-polorised photons could be post-selected. The weak measurement in C will therefore be negligibly small, as we have recently demonstrated in the lab for the single outer cycle considered here \cite{Salih2018b}. We can say that the photon was not in C during the first outer cycle. It was not in C during the second outer cycle. Therefore it was never in C. This generalises straightforwardly to setups with any larger number of inner and outer interferometers.

Interestingly, even for the paradoxical case arising from the pre-selected state being at the beginning of the first outer cycle and the post-selected state being at the end of the final outer cycle, a weak measurement in arm C approaches zero for a polarisation rotation by HWP1's $<<1$ (which as we will see is the case for related counterfactual communication and computation protocols ideally implemented). This is because the numerator of the weak value $\frac{\left\langle \Phi \right| O \left| \Psi \right\rangle}{\left\langle \Phi | \Psi \right\rangle}$ is then upper bounded by (the sine of) the small polarisation-rotation angle squared.

We reproduce the paradox and its resolution in the Appendix, from a time-symmetric consistent histories perspective. The fact that the paradox persists from another viewpoint specifically intended to unpick quantum paradoxes, is an indication of its reach. The paradox arises within an inconsistent framework, whereby probabilities assigned to series of events leading to mutually exclusive outcomes do not add up to unity---because of {\it after-the-fact} interference. As a result, if no consistent framework can be found for a given pre- and post-selection then any conclusions about the past of a particle drawn from corresponding weak values should be red-flagged as unreliable. Importantly, we were able to find a sequence of consistent frameworks to meaningfully ask the question of whether the photon was in any of the arms C. It was not.

Now, the setup in Fig. \ref{fig:Paradox} represents an instance of Salih et al.'s protocol for counterfactual communication of classical information \cite{Salih2013} with one of the communicating parties (Alice) on the left in control of all optical elements except mirrors $MR_B$, which are controlled by the other communicating party (Bob) on the right. The fact that the setup of Fig. \ref{fig:Paradox} employs a small number of nested interferometers means that there's photon loss as well as communication error, both of which can be eliminated in the limit of a large number of outer and inner interferometers. Further, communication error can be eliminated while employing a small number of cycles by introducing suitable loss-balancing attenuation in the outer interferometers. 

Bob then communicates bit ``0" by not blocking the optical path to Alice, arms C, as in the case discussed so far in Fig. \ref{fig:Paradox}. Provided the photon is not lost, detector $D_0$ clicks indicating a ``0". On the other hand, Bob communicates bit ``1" by blocking the optical path to Alice, with a photon detector for example. Provided the photon is not lost to Bob, detector $D_1$ instead clicks indicating a ``1". Here, the absence of interference in the inner interferometers allows the action of successive $HWP1$'s to rotate Alice's photon from H to V, leading to a $D_1$ click. Counterfactuality, which means that given a $D_0$ or $D_1$ click, Alice's photon has not travelled to Bob on the right, is not in question for the case of Bob blocking, since there is no dispute that any photon entering the channel would have been lost to Bob's blocking device. It is the case of Bob not blocking the channel where counterfactuality has been questioned, even though common sense tells us it must be the case, since any photon that enters the inner interferometers would necessarily be lost to detector $D_A$ through the action of successive $HWP2$'s. The demonstration above that the photon has not been to Bob for the case of Bob not blocking the channel therefore amounts, by the linearity of quantum mechanics, to a demonstration of counterfactuality not only of Salih et al.'s protocol for counterfactual communication of classical information\cite{Salih2013}, but also for the Salih14 generalised protocol for counterfactual communication of quantum information\cite{Salih2016,*Salih2014b}, where Bob implements a superposition of blocking and not blocking the channel, the focus of the rest of the paper.

\section*{Universal Exchange-Free Quantum Computation}
The exchange-free CNOT gate is the key primitive for the protocol for the counterfactual disembodied transport of an unknown qubit, counterportation, that was first proposed in \cite{Salih2016,*Salih2014b} as a generalisation of \cite{Salih2013}, drawing on ideas in \cite{Elitzur1993,Misra1977,Kwiat1999,Hosten2006,Noh2009,Nielsen2010}. The gate itself was the earliest counterfactual gate to be proposed, 2-qubit or otherwise, and allows universal exchange-free quantum computation as a new computing paradigm. The definition of counterfactuality, with the target qubit of the gate at one subsystem, Alice, and the control qubit at another, Bob, is that the two spatially-separated subsystems exchange no particles during the computation. In contrast to reference \cite{Salih2016,*Salih2014b}, we use the circular polarisation basis as it ties better with our experimental proposal for Bob's qubit. We provide a more succinct derivation in what follows before showing how to incorporate an improvement based on recent work by Aharonov and Vaidman, so as to establish broad agreement on counterfactuality. Consider a right-circular polarised, R, photon entering the top chained quantum Zeno effect module CQZE1 in Fig. \ref{fig:CQCNOT}C. As mentioned previously, the setup of Fig. \ref{fig:Paradox} is equivalent to two outer cycles here, each containing two inner cycles, except that in the present setup of Fig. \ref{fig:CQCNOT}C Bob implements a quantum superposition of blocking, which represents bit ``1", and not blocking the channel to Alice, bit ``0". It should become clear shortly what is meant by inner and outer cycles.

Note that we start with a separable product state of Alice's photon and Bob's atom. The action of the detectors helps to vividly explain counterfactuality, but is not necessary in obtaining the final state. This is because their measurements can be deferred, by the deferred measurement principle \cite{Nielsen2010}, to the end as that part of the quantum state no longer evolves. The overall process can therefore be taken to be unitary.

Switchable mirror SM1 is first switched off to allow the photon into the outer interferometer, before being switched on again. Switchable polarisation rotator SPR1, whose action is described by $\left| \text{R} \right\rangle \to \cos \frac{\pi}{2M} \left| \text{R} \right\rangle + \sin \frac{\pi}{2M} \left| \text{L} \right\rangle$, and $\left| \text{L} \right\rangle \to \cos \frac{\pi}{2M} \left| \text{L} \right\rangle - \sin \frac{\pi}{2M} \left| \text{R} \right\rangle$, rotates the photon's polarisation from R by a small angle $\frac{\pi}{2M}$. Polarising beam-splitter PBS2 passes the R part towards the bottom mirror while reflecting the small L part towards the inner interferometer. Switchable mirror SM2 is then switched off to allow the L part into the inner interferometer, before being switched on again. Switchable polarisation rotator SPR2 rotates the L part by a small angle $\frac{\pi}{2N}$. Polarising beam-splitter PBS3 then reflects the L part towards the top mirror while passing the small R part towards Bob, who is implementing a superposition, $\alpha \left| \text{0} \right\rangle + \beta \left| \text{1} \right\rangle$, of reflecting back any photon, and blocking the channel, respectively. More precisely, inside the inner interferometer, and given the photon is not lost to Bob's detector $D_B$,
\begin{align}
\left| \text{L} \right\rangle_{inner} (\alpha \left| \text{0} \right\rangle + \beta \left| \text{1} \right\rangle) \to \alpha (\cos \frac{\pi}{2N} \left| \text{L} \right\rangle_{inner} - \sin \frac{\pi}{2N} \left| \text{R} \right\rangle_{inner})\left| \text{0} \right\rangle + \beta \cos \frac{\pi}{2N} \left| \text{L} \right\rangle_{inner} \left| \text{1} \right\rangle
\end{align}

The R and L components of the photonic superposition are brought back together by PBS3 on the way towards SM2. This represents one inner cycle. After N such cycles we have,
\begin{align}
\left| \text{L} \right\rangle_{inner} (\alpha \left| \text{0} \right\rangle + \beta \left| \text{1} \right\rangle) \to - \alpha \left| \text{R} \right\rangle_{inner} \left| \text{0} \right\rangle + \beta {\cos}^N \frac{\pi}{2N} \left| \text{L} \right\rangle_{inner} \left| \text{1} \right\rangle
\end{align}

Switchable mirror SM2 is then switched off to let the photonic component inside the inner interferometer out. Since for large N, ${\cos}^N \frac{\pi}{2N}$ approaches 1, we have,
\begin{align}
\left| \text{L} \right\rangle_{inner} (\alpha \left| \text{0} \right\rangle + \beta \left| \text{1} \right\rangle) \to - \alpha \left| \text{R} \right\rangle_{inner} \left| \text{0} \right\rangle + \beta \left| \text{L} \right\rangle_{inner} \left| \text{1} \right\rangle
\end{align}

where the R-polarised part proceeds through PBS2 towards detector $D_A$, and the L part is reflected towards SM1. Importantly, since it is this R part of the superposition that has travelled to Bob, detector $D_A$ not clicking ensures that the photon has not been to Bob, which corresponds to the case of Bob's object not blocking. Whereas for the case of Bob's object blocking, detector $D_B$ not clicking ensures that the photon has not been to Bob.

Now that we know how a photonic component entering the inner interferometers evolves, we can look at the first outer cycle as a whole. Starting with the photon at SM1 travelling downwards we have, assuming the photon is not lost to Alice's detector $D_A$ or to Bob's $D_B$,
\begin{align}
\left| \text{R} \right\rangle (\alpha \left| \text{0} \right\rangle + \beta \left| \text{1} \right\rangle) \to \alpha \cos \frac{\pi}{2M} \left| \text{R} \right\rangle \left| \text{0} \right\rangle + \beta (\cos \frac{\pi}{2M} \left| \text{R} \right\rangle + \sin \frac{\pi}{2M} \left| \text{L} \right\rangle) \left| \text{1} \right\rangle
\end{align}

This represents one outer cycle containing N inner cycles. The photonic superposition has now been brought back together by PBS2 on its way to SM1. After M such cycles we have,
\begin{align}
\left| \text{R} \right\rangle (\alpha \left| \text{0} \right\rangle + \beta \left| \text{1} \right\rangle) \to \alpha {\cos}^M \frac{\pi}{2M} \left| \text{R} \right\rangle \left| \text{0} \right\rangle + \beta \left| \text{L} \right\rangle \left| \text{1} \right\rangle
\end{align}

Since for large M the term ${\cos}^M \frac{\pi}{2M}$ approaches 1, and with $M<<N$, which we will justify in the Methods and Results section when addressing losses, we have,
\begin{align}
\left| \text{R} \right\rangle (\alpha \left| \text{0} \right\rangle + \beta \left| \text{1} \right\rangle) \to \alpha \left| \text{R} \right\rangle \left| \text{0} \right\rangle + \beta \left| \text{L} \right\rangle \left| \text{1} \right\rangle
\end{align}

Switchable mirror SM1 is now switched off to let the photon out. Crucially, this last equation describes the action of a CNOT gate with Bob's trapped atom as the control qubit, acting on Alice's R-polarised photon, target qubit. But we want to allow Alice to input a superposition of R and L, hence the two CQZE modules in Fig. \ref{fig:CQCNOT}. Alice sends her photon into PBS1, which passes the R component towards the CQZE1 module as before, while reflecting any L component towards the bottom CQZE2 module where we have, 
\begin{align}
\left| \text{L} \right\rangle (\alpha \left| \text{0} \right\rangle + \beta \left| \text{1} \right\rangle) \to \alpha \left| \text{L} \right\rangle \left| \text{0} \right\rangle + \beta \left| \text{R} \right\rangle \left| \text{1} \right\rangle
\end{align}

The function of Pockels cell PC before CQZE2 is to flip the polarisation of any incoming L photon, as well as flipping the polarisation of any photon exiting CQZE2. This means that the two CQZE modules are identical. (In fact it might be more practical for the same CQZE module to double as CQZE1 and CQZE2. This can be achieved by using a PBS1 that splits the R and L photonic components into two parallel modes that are then fed to the same CQZE module.)

By means of optical circulators OC1 and OC2, the outputs of the two CQZE modules are directed towards 50-50 beam-splitter BS, where they are added together correctly at Port1. At Port2, however, the output from CQZE2 acquires a negative phase, by the action of BS, as it is added to the output from CQZE1. This has an effect, given the photon is found at Port2, equivalent to applying a Z-gate to the polarisation of Alice's photon before being initially sent into PBS1. This can be corrected by applying Z transformations by Alice and Bob on their respective qubits at the end, but entails Alice sending Bob one bit of information exchange-free. This can easily be done using another exchange-free CNOT with a classical control. (Interestingly, as we will see, no such exchange-free classical communication is needed in the counterportation protocol.) Computationally, the CNOT gate is universal, meaning that any quantum circuit can be efficiently implemented using only CNOTs and single-qubit operations \cite{Nielsen2010}. Consequently, any exchange-free quantum computation---{\it defined as a quantum computation over spatially-separated subsystems in the absence of particle exchange}---is possible using the exchange-free CNOT gate and single-qubit operations. This completes the gate's description.

We have recently demonstrated in the lab that the laws of physics do not prohibit counterfactual communication\cite{Salih2018b}. We were happy to lose most photons during communication, so long as counterfactuality was unequivocally established, which we were able to do by employing a single outer cycle. For the multiple outer-cycles considered here, where the probability of photon loss can be made arbitrarily close to zero, we have already shown by means of a new quantum paradox that a nonzero weak measurement does not necessarily mean the photon was at Bob. Further, we have shown through the resolution of the paradox that the photon was in fact never at Bob, regardless of the number cycles. It would be nice, however, if the result of a weak measurement at Bob is zero given the pre- and post-selected states are at the beginning and at end of the protocol, respectively. The recently proposed modification by Aharonov and Vaidman\cite{Aharonov2019Modificationtraces} of the version of Salih et al.'s counterfactual communication \cite{Salih2013} that does not use polarisation, while not passing the consistent-histories test, as shown in \cite{Salih2018b}, does make the result of a weak measurement at Bob zero as a first order approximation. Importantly, the analysis in the Appendix below shows that the use of the polarisation degree of freedom allows counterfactuality to be established from a consistent-histories viewpoint. It is straightforward to check that this still holds if the inner-cycle sequence is repeated, which is what Aharonov and Vaidman's modification entails.

Here's how to implement Aharonov and Vaidman's modification in the exchange-free CNOT gate. Take the Nth inner cycle, which was previously the last inner cycle during a given outer cycle. After applying SPR2 inside the inner interferometer for the Nth cycle, Alice makes a measurement by blocking the entrance to the channel leading to Bob. (She may alternatively flip the polarisation and use a PBS to direct the photonic component away from Bob.) And instead of switching SM2 off, it is kept turned on for a duration corresponding to N more inner cycles, after which SM2 is switched off as before. One has to compensate for the added time by means of optical delays. The idea here is that, for the case of Bob not blocking, any lingering V component inside the inner interferometer after N inner cycles (because of weak measurement or otherwise) will be rotated towards H over the extra N inner cycles. This has the effect that, at least as a first order approximation, any weak measurement on the photon in the channel leading to Bob, including one of polarisation, will be vanishingly small. In the language of the TSVF, the forward state and the backward state do not overlap anywhere in the channel.

%In fact one can make the result of a weak measurement at Bob arbitrarily small. The way to do this is by repeating the same trick again, employing N more inner cycles---namely by blocking the entrance to the channel leading to Bob once more after the 2Nth application of SPR2, or else directing the photonic component away, then keeping SM2 on for a duration corresponding to N more inner cycles---and if one wishes, this can be repeated again and again.

\section*{Extended Counterportation}
With the quantum circuits in Fig. \ref{fig:CQCNOT}A and B from reference \cite{Salih2016,*Salih2014b} in mind, Alice sends her R-polarised photon from the left towards PBS1, as shown in Fig. \ref{fig:CQCNOT}C, where the photon proceeds towards CQZE1 at the top, whose action corresponds to the first CNOT in Fig. \ref{fig:CQCNOT}A. With Bob implementing his qubit as a superposition of reflecting and blocking, $ \alpha \left| \text{0} \right\rangle + \beta \left| \text{1} \right\rangle$, Alice's photon emerges back maximally entangled with Bob's qubit. Optical circulator OC, and then switchable mirror SM0 briefly turned on, reflect the photon into Port0, where a Hadamard transformation is applied to its polarisation. A Hadamard transformation is also applied to Bob's qubit by means of suitable laser pulses. Alice's photon is then fed again into PBS1 from the left. The R-polarised component incident on PBS1 proceeds towards CQZE1 as before, while the L-polarised component is reflected by PBS1 towards Pockels cell PC, which flips its polarisation to R, before entering CQZE2. The component that eventually emerges from CQZE2 will have its polarisation flipped again by PC on its way back before being directed by optical circulator OC2 towards beam-splitter BS. This photon component combines with the photon component emerging from CQZE1 and directed by OC1 towards beam-splitter BS. This corresponds to the second CNOT in Fig. \ref{fig:CQCNOT}A and B. A Hadamard transformation is then applied to Bob's qubit. Similarly, Hadamard transformations are applied to the polarisation of the photon components in ports 1 and 2, before a NOT transformation is applied to the polarisation of the photon component at port2 in order to compensate for the effective phase-flip introduced by beam-splitter BS, as shown in Fig. \ref{fig:CQCNOT}B. The photonic components at both ports now have identical polarisation. Found in either of Alice's ports, which happens with unit probability in the limit of $N>>M>>1$ and perfect implementation, where M and N are the number of outer and inner CQZE cycles respectively, polarisation is the desired $\alpha \left| \text{R} \right\rangle + \beta \left| \text{L} \right\rangle$.

By way of extension, consider Bob in possession of n qubits, collectively comprising an unknown, possibly entangled state, and consider Alice in possession of n, initially separable R-polarised photons. Using the counterprotation protocol above, each of Bob's qubits can be counterported to one of Alice's photons. By the linearity of quantum mechanics, Bob's entire unknown quantum state is thus counterported to Alice. Further, each of the qubits comprising the state to be counterported can be with a different Bob, and each of the photons can likewise be with a different Alice, allowing counterportation to be decentralised. Lastly, mixed states can be counterported using the protocol even when a small number of cycles is employed. For this, photon loss for the case of Bob blocking the channel needs to match that for the case of Bob not blocking the channel. This can be achieved, as explained in the next section, by introducing appropriate attenuation that balances photon loss.

\section*{Methods and Results}
Bob needs to implement a superposition of reflecting Alice's photon, bit ``0'', and blocking it, bit ``1''. There are various ways to go about this, including cavity optomechanics\cite{Aspelmeyer2014CavityOptomechanics} and quantum dots\cite{Lodahl2015InterfacingNanostructures,Guo}. However, recent breakthroughs in trapped atoms inside optical cavities\cite{Reiserer2015Cavity-basedPhotons}, including the experimental demonstration of light-matter quantum logic gates\cite{Tiecke2014NanophotonicAtom,Reiserer2014a}, make trapped atoms an attractive choice.

Here, a single $^{87}$Rb atom trapped inside a high-finesse optical resonator by means of a three-dimensional optical lattice constitutes Bob's qubit\cite{Reiserer2014a,Reiserer2014}. Depending on which of its two internal ground states the $^{87}$Rb atom is in, a resonant R-polarised photon impinging on the cavity from the left in Fig. \ref{fig:CQCNOT}C will either be reflected as a result of strong coupling, or otherwise enter the cavity on its way towards detector $D_B$. Unlike references \cite{Reiserer2014a,Reiserer2014}, for our purposes here, the cavity needs to, first, support the two optical modes shown in Fig. \ref{fig:CQCNOT}C (or else support two parallel optical modes impinging on the cavity from the same side, as in \cite{Hacker2016AResonator}, which ties in with the earlier suggestion to use a single CQZE module with two parallel optical modes). And second, it needs to have (equal) mirror reflectivities such that a photon entering the cavity exits towards detector $D_B$, similar to \cite{Mucke2010}. By placing the $^{87}$Rb in a superposition of its two ground states, by means of Raman transitions applied through a pair of Raman lasers, Bob implements the desired superposition of reflecting Alice's photon back and blocking it. Note that coherence time for such a system is on the order of 0.1 millisecond\cite{Reiserer2014}, with longer coherence times possible. Therefore, if the protocol is completed within a timescale on the order of microseconds or tens of microseconds, which is lower-bounded by the switching speed of switchable optical devices, on the order of nanoseconds, then decoherence effects can be ignored. (There are experimental tricks for ensuring the correct number of cycles without having to use switchable optical elements, as in, for instance, the recent experimental implementation of Salih et al.'s counterfactual communication by Cao et al.\cite{Cao2017}.)

We numerically simulate counterfactual disembodied transport by means of recursive relations based on the ones in \cite{Salih2013}, which track the evolution of Alice's photon from one cycle to the next depending on Bob's bit choice. 

Interestingly, by adding suitable attenuation that balances loss, we can eliminate any dependency of fidelity on the number of inner and outer cycles, given perfect implementation. First, considering the CQZE's modules in Fig. \ref{fig:CQCNOT}C, the R component in each outer cycle that goes through the optical delay at the bottom is attenuated, following reference \cite{Salih2020a}, by a factor of $({\cos} \frac{\pi}{2N})^N$, in order to balance loss in the L component caused by Bob blocking. This balanced loss allows us to intuitively understand where the earlier mentioned condition $M<<N$ for making loss approach zero for the case of Bob blocking, comes from. Importantly, the factor in question is raised to the power M by the time all outer cycles are completed. We can write $({\cos} \frac{\pi}{2N})^{N\times M}$ as $({\cos} \frac{\pi}{2N})^{N^2\times M/N}$, which as N approaches infinity tends to $1/e^\frac{\pi^2}{8}$ for $M=N$, and tends to 1 for $M<<N$. Second, in the final outer cycle, between SM2 and PBS2, we introduce an attenuation of the L component exiting the inner cycles by a factor of $({\cos} \frac{\pi}{2M})^M$, in order to balance loss in the R component caused by Bob not blocking. The success probability of counterportation, or equivalently for reasons upcoming in the Discussion, the probability of traversing a local wormhole,
\begin{align}
P_\mathcal{LWH}=\cos^{2M}(\theta_M) cos^{2MN}(\theta_N)
\end{align}
with $P_\mathcal{LWH}\to1$ for $N>>M>>1$, where $\theta_M=\frac{\pi}{2M}$ and $\theta_N=\frac{\pi}{2N}$ are the polarisation rotations per outer- and inner-cycle respectively, and M and N are the number of outer- and inner-cycles respectively. Note that this equation is exact. Remarkably, counterportation fidelity $F=1$, irrespective of M and N. Further, since loss is invariant across input states, the protocol can take mixed states as input. Added attenuation, however, is less effective when device imperfections are high, which is the case for current technology, and is therefore not implemented in the simulations below.

We now account for two types of imperfections corresponding to Bob reflecting and not reflecting the photon back. First, imperfections obstructing the communication channel. Second, Bob's cavity failing to reflect the photon back, which for the experimental setup in reference \cite{Reiserer2014} happened with probability 34(2)\%. This is caused by scattering or absorption within the cavity. However, dramatically reduced loss is expected for next-generation cavities with increased atom-cavity coupling strength. In our numerical simulation we combine these two types of imperfections into one coefficient associated with Bob reflecting the photon. We assume that the probability of imperfections obstructing the channel to be small enough to lie within the uncertainty for Bob's object failing to reflect, and to be ignored. For the case of Bob blocking the channel, we account for imperfect optical mode matching, that is imperfect transverse overlap between the free-space mode of the photon and the cavity mode. For the setup in reference \cite{Reiserer2014}, this happened with probability 8(3)\%, and resulted in the photon being reflected back when it should not. This is also expected to improve with next-generation cavities.

Fidelity of counterfactual transport is plotted in Fig. \ref{fig:Fidelity}, averaged over 100 evenly distributed qubits on the Bloch sphere, for a number of outer cycles up to 10, and a number of inner cycles up to 10. Fidelity $F=\left\langle \psi_{in} \right| \rho_{out}  \left| \psi_{in} \right\rangle$, where $\psi_{in}$ is the state to be counterported from Bob, and $\rho_{out}$ is the density matrix of the counterported state to Alice. Here, an error coefficient of 34\% associated with Bob reflecting the photon is assumed, along with an error coefficient of 8\% associated with Bob blocking. Fidelity, even for the minimum number of CQZE cycles, 4, is above the average classical fidelity limit of 2/3 \cite{Popescu1995}. In this case, efficiency is only 0.1\%, but improves towards unity by increasing the number of cycles and reducing device imperfections. For example, for 20 outer cycles and 75 inner cycles, and error coefficients 4\% and 1\%, efficiency is above 50\%, with fidelity close to 99\%.

%Meanwhile, with Aharonov and Vaidman's modification adapted for our setup, for a total number of cycles of 16, and with the above error coefficients of 34\% and 8\%, efficiency and fidelity are above 4\% and 75\% respectively.

The original concept of counterportation, which was proposed in 2014 \cite{Salih2016,*Salih2014b}, has lead to a subsequent proposal more than a year later \cite{Li2015DirectState} that was based, as Vaidman noted \cite{Vaidman2016}, on the same basic concept we had proposed. Importantly, given realistic imperfections, the fidelity of their experimental proposal falls below the 2/3 limit, even for thousands of protocol cycles. In fact, one would have to employ an impractical tens of thousands of cycles, as can be seen from their Figure 4. By contrast, in our proposed realisation, for 2 outer cycles and 3 inner cycles for instance, which amounts to 12 protocol cycles in total as the exchange-free CNOT is twice instantiated---and with Bob's realistic device failing to reflect 34\% of the time and failing to block 8\% of the time---fidelity is already close to 75\%. This is promising for a near-future experimental demonstration.

\section*{Discussion}
Our universal exchange-free quantum computation, not to be confused with its quaint predecessor that allows a computer to perform a computation without actually ``running'' \cite{Jozsa2001,Hosten2006}, generalises exchange-free communication---all of which have been inspired by, and intuitively explained in terms of ``interaction-free'' measurement \cite{Elitzur1993}, and the Zeno effect \cite{Misra1977}. In interaction-free measurement the presence of a measuring device, or the ``bomb'' in Elitzur and Vaidman's thought experiment, can sometimes be inferred without any particle triggering it. Whereas in the Zeno effect, which can be used to boost the efficiency of interaction-free measurement, repeated measurement of a quantum state inhibits its evolution, leaving it unchanged (the proverbial watched kettle that does not boil). 

Let's consider detectors $D_A$ and $D_B$ in Fig. \ref{fig:CQCNOT}C. By the deferred measurement principle\cite{Nielsen2010}, which states that any part of a quantum system that has stopped evolving can be measured straightaway or at a later time, we can imagine detectors $D_A$ and $D_B$ being placed far away such that neither performs any measurement before the photon could exit the protocol. At the end of the protocol, the photon is in Port1 and Port2 with probability amplitude approaching unity in the limit of $N>>M>>1$ and perfect implementation, in the desired polarisation state of Bob's original qubit. No reference to either interaction-free measurement or the Zeno effect is therefore required, showing that neither is a necessary condition for counterfactuality.

We make the assumption that optical communication is explainable by one or more of the following: 1) detectable photons traversing the channel between the two communicating parties; 2) measurements carried out in between an initial state and a final state; and 3) an underlying physical state objectively existing prior to measurement.

When exchange-free communication is cast in terms of interaction-free measurement and the Zeno effect, repeated measurements appear to play a key role in information transfer---with quantum collapse due to measurement, in the words of the authors of the PBR paper \cite{Pusey2012OnState}, a problematic and poorly defined physical process. By contrast, the correct counterported state in the present protocol can be post-selected with probability approaching unity in the absence of such measurements, and moreover, with the post-selection not involving any measurement of information-carrying qubits. Thus the updating of the experimenter's subjective knowledge based on measurement outcomes plays no part in information transfer.

By presenting a deterministic scheme for exchange-free communication that does not involve any measurements in between an initial state and a final state, what we are left with as a possible carrier of quantum information---in the absence of particle exchange---is an underlying physical state that exists objectively prior to measurement. But Colbeck and Renner have showed using quite an involved argument that, under the assumption of freedom of choice of experimental settings, the quantum state $\Psi$ is complete, meaning that its experimental predictions cannot be improved upon by any future extension of quantum theory \cite{Colbeck2011}. This implies the completeness of the underlying physical state $\Lambda$, assuming its existence, since $\Lambda$ can always include $\Psi$. Yet, the existence of an objective, complete $\Lambda$ follows directly from the fact that communicating an unknown qubit in counterportation, which violates the classical fidelity limit for disembodied transport exchange-free, takes place deterministically prior to any measurement, the outcome of which, namely finding the photon at Alice, is also deterministic. The authors of the PBR paper, in their seminal investigation of the reality of the quantum state $\Psi$ have shown that given a few reasonable assumptions \cite{Pusey2012OnState}, which Colbeck and Renner have subsequently replaced with only that of freedom of choice of experimental settings \cite{Colbeck2017AState}, the quantum state $\Psi$ is real, in the Harrigan and Spekkens sense \cite{Spekkens} that multiple quantum states cannot correspond to the same underlying physical state $\Lambda$---if such a state exists. What we have provided here is evidence that an underlying physical state does exist, and is what has counterported an unknown, possibly entangled quantum state across space.

In counterportation, Bob communicates information by enabling single-photon self-interference to take place either in the inner interferometers but not the outer when communicating a ``0'', or in the outer interferometers but not the inner when communicating a ``1'', or as we have shown, in a linear combination of the two scenarios when communicating a qubit. In the second quantisation, as we are reminded in \cite{Susskind2016}, the path superposition of a single photon constitutes entanglement across space, as illustrated in Fig.\ref{fig:SpatialEntanglement}. This is manipulated in the two self-interference scenarios of counterprotation just mentioned, to gradually entangle then gradually unentangle the two parties' qubits.

Tantalisingly, a strong version of the recently proposed ER=EPR conjecture \cite{Susskind,Susskind2016} asserts that any pair of entangled qubits across space, referred to as EPR after Einstein, Podolsky, and Rosen \cite{PhysRev.47.777}, is spatially connected through some sort of Einstein-Rosen (ER) bridge or wormhole \cite{PhysRev.48.73} no matter how far the qubits from one another.

We will now outline as an explanatory framework for counterportation a general yet precise characterisation---starting from the constructor theory of information \cite{Marletto2014}---of what will be called a local ER bridge or local wormhole for reasons that should become clear. A constructor is a substrate that can carry out a {\it task} on a subsystem, another substrate, with arbitrarily high accuracy, while retaining the ability to do so again. If such a constructor could exist under the laws of physics then the corresponding task is possible, otherwise it is impossible. Particularly important then is the question of which tasks are possible and which are not. This counter-intuitive shift in emphasis from what will happen to what could happen, has recently lead to surprising new insights, including a remarkable proposal for experimentally testing the quantisation of gravity \cite{Marletto2017b}. 

From an information viewpoint, constructor theory makes reference to {\it information media}, whose defining task is the possibility of copying information. This is the hallmark of classical-like systems, where information is {\it interoperable}: It can be copied from one medium, or substrate, to another while retaining all its properties as information, making it substrate-independent. The theory also makes reference to {\it super-information media}, whose defining task by contrast is the impossibility of copying or ``cloning''. Super-information media are thus quantum-like. Another constructor-theoretic principle, especially important for our purposes here is that of {\it locality}. The Principle of locality asserts the possibility of a description whereby any task carried out on some subsystem, out of multiple subsystems, can only change that subsystem and no other.

This puts us in a position to define a new constructor, namely a universal exchange-free 2-qubit gate. Based on this constructor: {\it The task of performing any quantum computation over spatially-separated subsystems is possible without particle exchange}.

Alice and Bob's states, which we assume are qubits, constitute one sector of the system. By the principle of locality, the constructor, of which the universal exchange-free CNOT gate is an example, which we recall can entangle Alice and Bob's qubits, requires another sector to mediate any entangling action over the two spatially separated qubits---otherwise acting on either qubit separately cannot change the other. Note that given the constructor, one need not make any initial assumptions about the dynamics governing its mediating sector. Yet, strikingly, Marletto and Vedral, invoking the interoperability principle, have shown that any medium capable of entangling two subsystems is necessarily a super-information medium, i.e. characterised by such properties as no-cloning \cite{Marletto2017a}.

The repertoire of exchange-free quantum computation, mediated by the constructor's super-information medium, includes counterportation---which makes possible the defining task of a traversable wormhole, namely rendering space traversable in the absence of any journey across space observable by an outside observer, in other words {\it rendering space traversable disjunctly}. We therefore identify the super-information medium as a local wormhole. The descriptor ``local'' is in contrast to classical ER bridges (which arise as solutions of general relativity) since Alice and Bob are spatially separated rather than space-like separated. 

Dynamically, while spatial superpositions of a single particle can entangle far-apart regions of space through local propagation as illustrated in Fig. \ref{fig:SpatialEntanglement}, carefully choreographed self-interference scenarios can, as shown in the Appendix, employ such spatial entanglements to entangle other degrees of freedom across the two regions exchange-free, in our example-constructor photon polarisation and cavity reflectively. This is in contrast to the special case of a classical Bob \cite{Salih2018}, where the need for a super-information medium is not as obvious \cite{Gisin2013}.

In our conception of ER=EPR, {\it a local wormhole is a physical system characterised by the no-cloning of information that mediates the action of an exchange-free quantum computer: Locally induced spatial entanglements of a single particle, and corresponding self-interference scenarios allow entangling other degrees of freedom among spatially separated parties, without any particle observably crossing---rendering space disjunctly traversabile}.

A constructor-theoretic approach has thus enabled us to precisely characterise a local ER bridge in terms of a task/constructor pairing, namely the disjunct traversability of space by means of an exchange-free quantum computer, without assuming any specific quantum gravity dynamics. Explaining counterportation, EPR provides a dynamical account of how a local wormhole mediates qubit exchange-free interaction.

A recent result \cite{Aharonov2020} has beautifully illustrated the local conservation of modular angular momentum of exchange-free communication. It has showed, using a simplified setup, that a local change of modular angular momentum in the form of a $\pi$ phase-rotation of Bob's qubit is mirrored by a corresponding local change of modular angular momentum of Alice's photon, of exactly $\hbar$. But as we've shown in \cite{Salih2020b}, such a phase flip can be used to construct a universal exchange-free controlled-Z gate. And since counterportation is implementable using two controlled-Z's, and by the symmetry of the counterportation protocol thus implemented, the net change of modular angular momentum in either qubit in such counterportation is zero. This means that information transfer is not accounted for by a net change of modular angular momentum. Understanding conditional $\pi$ phase-rotation, and resulting transient entanglement, as a conditional  $\hbar$ local change of modular angular momentum in one qubit, with a corresponding conditional  $\hbar$ local change in modular angular momentum in the other, is nonetheless insightful. 

Conserved quantities in a quantum system are eigenstates of the system as a whole, which in this case is spatially entangled---hinting that in exchange-free operations spatial entanglement is the deeper explanation at work. Whereas one would have expected the super-information medium to obey conservation relations, the result in \cite{Aharonov2020} provides important confirmation.

While we envisage that future quantum-gravity considerations could help unravel the physics of local wormholes, the general yet precise characterisation given, already provides an explanatory framework for counterportation. Nonetheless, as geometrical conceptions rendering space disjunctly traversable \cite{Bronnikov2022} further advance, an experimental demonstration of counterportation---and therefore disjunct space traversability---would point not only to the existence in the lab of traversable wormholes, but conceivably of such geometry. The demonstration requires what is essentially a 2-qubit exchange-free quantum computer. By contrast to large-scale quantum computers that promise some remarkable speed-ups, the promise of exchange-free quantum computers of any scale is to make some seemingly impossible transformations possible by incorporating space fundamentally. Ultimately, we see exchange-free quantum computers coming into their own exploring the fundamental physics of the universe.

Accomplishing the mission of counterporting an object across space---thereby traversing a local wormhole, as if through an extra spatial dimension---comes down to locally induced spatial entanglements of a single particle, and corresponding self-interference scenarios at work. But as Richard Feynman was quick to point out, such interference ``has in it the heart of quantum mechanics.'' Even further, the uncovered phenomenon of counterportation provides a smoking gun for the existence of an underlying physical reality.

\newpage

\section*{Author Contribution}
This is a single-author paper.

\section*{Data Availability}
The datasets generated and analysed during the current study are available from the author on reasonable request.

\section*{Competing Interests}
The author declares that there are no competing interests.

\section*{Acknowledgments}
I thank Terry Rudolph, Will McCutcheon, Jonte Hance, Paul Skrzypczyk, Sandu Popescu, and John Rarity for helpful conversations; Robert Griffiths, Sam Braunstein, and Chiara Marletto for helpful email correspondence. This work was supported by the UK's Engineering and Physical Sciences Research Council (EP/PS10269/1). 

%Much has been said of buildings/institutions named after historical figures with ``complex legacies''. One such figure is H. H. Wills, a Bristolian philanthropist born in the 19th century into the Wills tobacco family, whose thriving business would have profited from slave-grown tobacco. More than a historical curiosity, such legacies are relevant to present-day inequalities.

\appendix
\section*{Appendices}
\subsection*{Local Wormhole from Single-Particle Spatial Entanglement}
We revisit the exchange-free CNOT gate from the section above, only now we explicitly keep track of path information using second quantisation. We define the qubit $\left| \text{1} \right\rangle_A$ as the photon on a path avoiding Bob, exiting through Alice's SM1 in Fig.\ref{fig:CQCNOT}C, and the qubit $\left| \text{0} \right\rangle_A$ as the photon not on that path. We also define the qubit $\left| \text{1} \right\rangle_{B'}$ as the photon on the path through Bob towards detector $D_A$, and the qubit $\left| \text{0} \right\rangle_{B'}$ as the photon not on that path; and the qubit $\left| \text{1} \right\rangle_{B''}$ as the photon on the path through Bob towards detector $D_B$, and the qubit $\left| \text{0} \right\rangle_{B''}$ as the photon not on that path. The action of the exchange-free CNOT gate, captured by equations 1 to 7, on Alice's R-polarised photon, for $N>>M>>1$, can be expressed as,

\begin{align}
\begin{split}
& \left| \text{1} \right\rangle_A \left| \text{0} \right\rangle_{B'} \left| \text{0} \right\rangle_{B''} \left| \text{R} \right\rangle (\alpha\left| \text{0} \right\rangle + \beta\left| \text{1} \right\rangle) \to \\ &
\alpha(\left| \text{1} \right\rangle_A \left| \text{0} \right\rangle_{B'} \left| \text{0} \right\rangle_{B''} \left| \text{R} \right\rangle \left| \text{Remaining} \right\rangle + \cancelto{0}{\left| \text{0} \right\rangle_A \left| \text{1} \right\rangle_{B'} \left| \text{0} \right\rangle_{B''} \left| \text{R} \right\rangle \left| \text{M leaks} \right\rangle})\left| \text{0} \right\rangle + \\ &
\beta(\left| \text{1} \right\rangle_A \left| \text{0} \right\rangle_{B'} \left| \text{0} \right\rangle_{B''} \left| \text{L} \right\rangle \left| \text{Remaining} \right\rangle + \cancelto{0}{\left| \text{0} \right\rangle_A \left| \text{0} \right\rangle_{B'} \left| \text{1} \right\rangle_{B''} \left| \text{R} \right\rangle \left| \text{M$\times$N leaks} \right\rangle})\left| \text{1} \right\rangle
\end{split}
\end{align}
\\
where $\left| \text{M leaks} \right\rangle$ sums all the photonic probability amplitudes that leak through Bob towards detector $D_A$ by means of self-interference in the inner cycles, for the case of Bob not blocking, denoted $\left| \text{0} \right\rangle$. These take place, as previously shown, one leak at the end of each of the M outer-cycles, repeatedly removing any polarisation rotation of the photonic component that has remained at Alice, leaving it unchanged at R as no interference takes place in the outer cycles. On the other hand, $\left| \text{M$\times$N leaks} \right\rangle$ sums all the photonic probability amplitudes that leak through Bob towards detector $D_B$, with no self-interference in the inner cycles, for the case of Bob blocking, denoted $\left| \text{1} \right\rangle$. These take place one leak at each of the M$\times$N inner cycles, allowing polarisation rotations of the photonic component that has remained at Alice to build up all the way to L as interference takes place in the outer cycles. $\left| \text{M leaks} \right\rangle$ and $\left| \text{M$\times$N leaks} \right\rangle$ can thus be considered as consisting of a series of time-bin states, with $\left| \text{Remaining} \right\rangle$ as a reference time-bin whose probability amplitude approaches unity in the limit of $N>>M>>1$. While, correspondingly, $\left| \text{M leaks} \right\rangle$ and $\left| \text{M$\times$N leaks} \right\rangle$ each approaches zero as individual leaks tend to zero, successful post-selection of $\left| \text{1} \right\rangle_A \left| \text{0} \right\rangle_{B'} \left| \text{0} \right\rangle_{B''}$ ensures that the photon has not been to Bob, even for the smallest possible M and N. (By means of suitable attenuation at Alice, the probability of photon loss for either leakage scenario is the same.) Yet, spatial entanglement between Alice and Bob has been created, and only destroyed in the final post-selection of $\left| \text{1} \right\rangle_A \left| \text{0} \right\rangle_{B'} \left| \text{0} \right\rangle_{B''}$. Remarkably, what has locally allowed polarisation to stay as $\left| \text{R} \right\rangle$ or rotate to $\left| \text{L} \right\rangle$, is respectively the spatial entanglement of the path denoted by $A$ with the path denoted by $B'$ on the one hand, and with the path denote by $B''$ on the other.

Locally induced by Bob, each spatial entanglement of Alice's single photon corresponds to a distinct scenario consisting of carefully choreographed photonic leakage from Alice to Bob. By means of more-frequent leakage steps, one leakage scenario allows self-interference to incrementally rotate Alice's polarisation along her path that never goes to Bob. By means of less-frequent leakage steps, interestingly, the other leakage scenario allows no such rotation buildup, as self-interference takes place elsewhere. Therefore given post-selection,
\begin{align}
\left| \text{1} \right\rangle_A \left| \text{0} \right\rangle_{B'} \left| \text{0} \right\rangle_{B''} \left| \text{R} \right\rangle (\alpha\left| \text{0} \right\rangle + \beta\left| \text{1} \right\rangle)
 \to \left| \text{1} \right\rangle_A \left| \text{0} \right\rangle_{B'} \left| \text{0} \right\rangle_{B''} (\alpha\left| \text{R} \right\rangle \left| \text{0} \right\rangle + \beta\left| \text{L} \right\rangle \left| \text{1} \right\rangle)
\end{align}

which sums up the action of the local wormhole of entangling the polarisation of Alice's photon with the reflectivity of Bob's object, as well as the fact that Alice's photon, having not been to Bob, has remained with her all the while.

\subsection*{Paradox and Resolution Using Consistent Histories}
\begin{figure}[ht]
\centering
\includegraphics[width=0.70\textwidth]{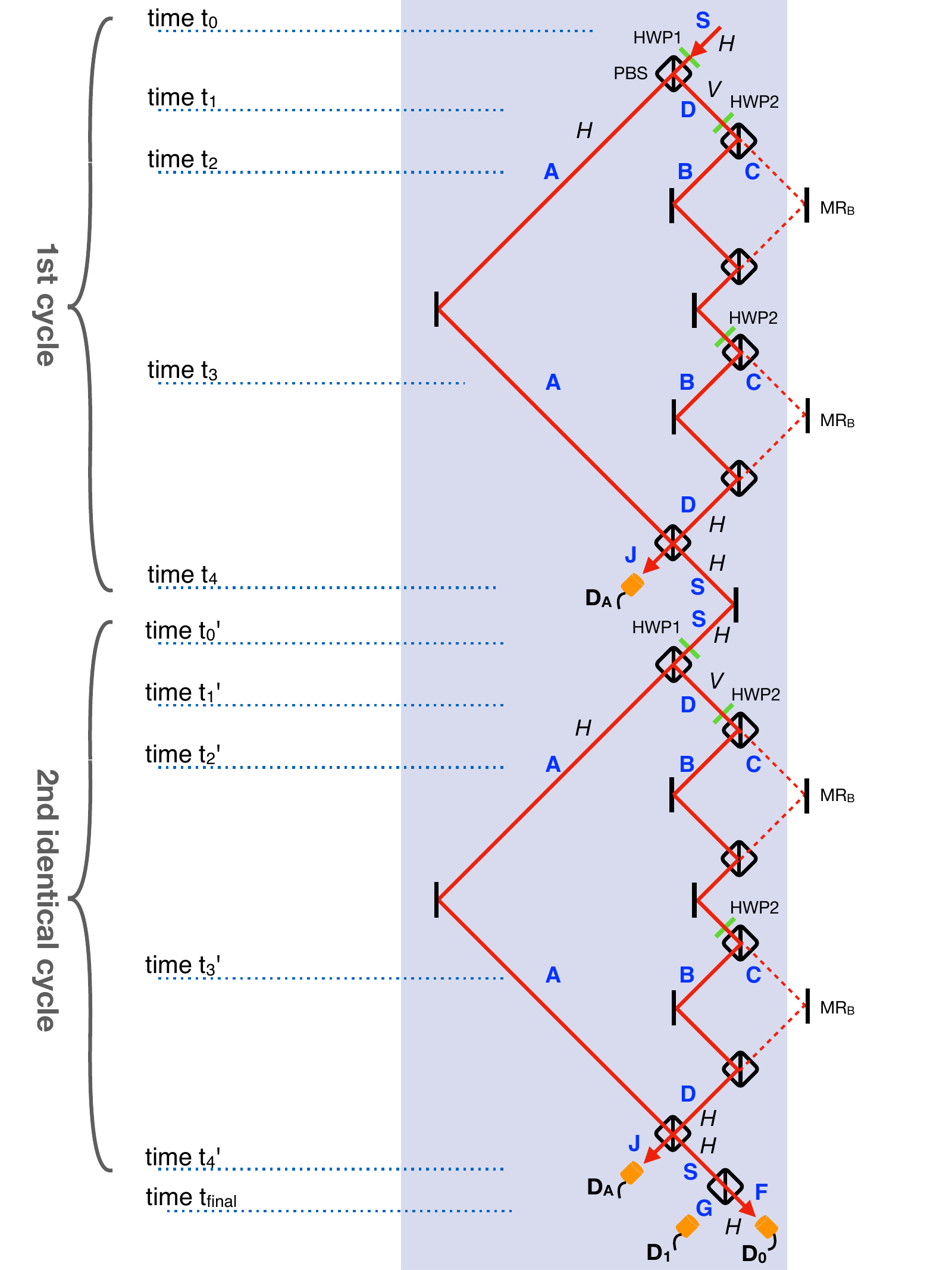}
\caption{\label{fig:Paradox}Two outer interferometers, nested within each are two inner interferometers. All beam-splitters are polarising, and all half-wave plates (HWP's) rotate polarisation by 45 degrees. Interferometer arms are labelled in blue. The photon starts at the top, H-polarised. The set-up is such that any photon entering the inner interferometers ends up at detector $D_A$. Given that the photon is not detected by either $D_A$, its evolution between times $t_0$ and $t_4$ is identical to its evolution between times $t'_0$ and $t'_4$. We want to know whether a photon detected at $D_0$ at the bottom was in any of the arms labeled C leading to the mirrors on the right---which common sense tells us should not be the case. However, consideration of weak measurements of the path observable at arms C can give paradoxical answers. Our resolution of the paradox, supported by recent weak-measurement {\it experimental results} \protect\cite{Salih2018b}, shows that the photon has not been to any of the right-hand-side mirrors. The deeper significance is that such a setup can be adapted, as explained in the text, for communication, and more generally, quantum computation across space without particle exchange.}
\end{figure}

\begin{figure}[ht]
\centering
\includegraphics[width=\linewidth]{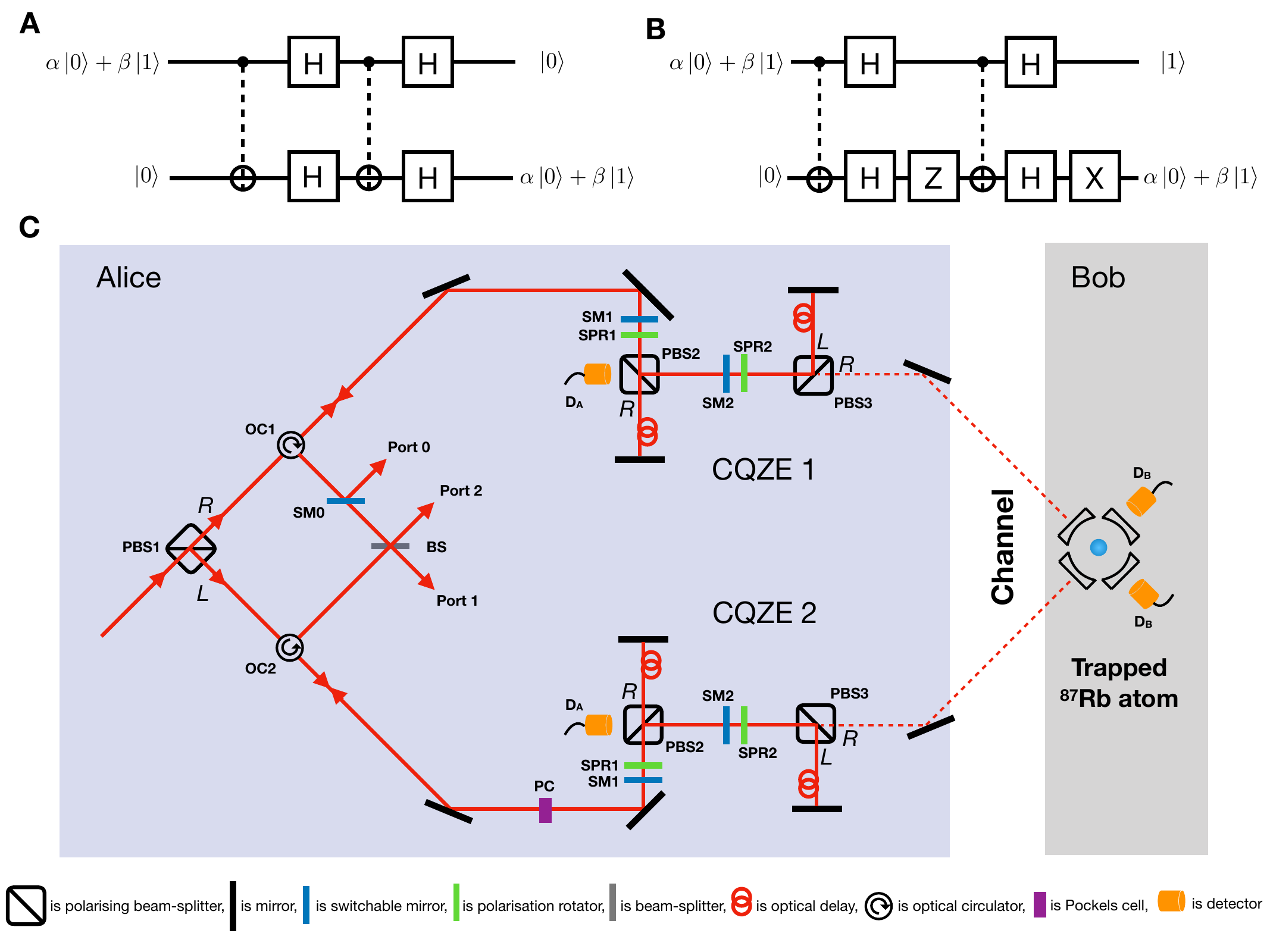}
\caption{\label{fig:CQCNOT}A) shows a circuit diagram for transporting Bob's qubit, $\alpha \left| \text{0} \right\rangle + \beta \left| \text{1} \right\rangle$, to Alice by means of two exchange-free CNOT gates and local operations. The purpose of the Hadamard gates is to keep the control qubits of the two CNOT gates on the same side, Bob's side. B) A similar circuit, except for the phase-flip Z-gate acting on Alice's target qubit before the second CNOT, which corresponds to finding the photon in Port2 in Fig. \ref{fig:CQCNOT}C, after the second application of the exchange-free CNOT gate. C) Our exchange-free CNOT gate. A single $^{87}$Rb atom trapped inside an optical resonator constitutes Bob's control qubit. Depending on which of two ground states the trapped atom is in, a resonant R-polarised photon impinging on the cavity from the left will either be reflected as a result of strong coupling, or else enter the cavity on its way towards detector $D_B$. CQZE stands for chained quantum Zeno effect. As we show in the text, Alice's exiting photonic qubit, the target, has provably never crossed the channel to Bob. This CNOT gate allows universal exchange-free quantum computation, including counterportation.}
\end{figure}

\begin{figure}[ht]
\centering
\includegraphics[width=\linewidth]{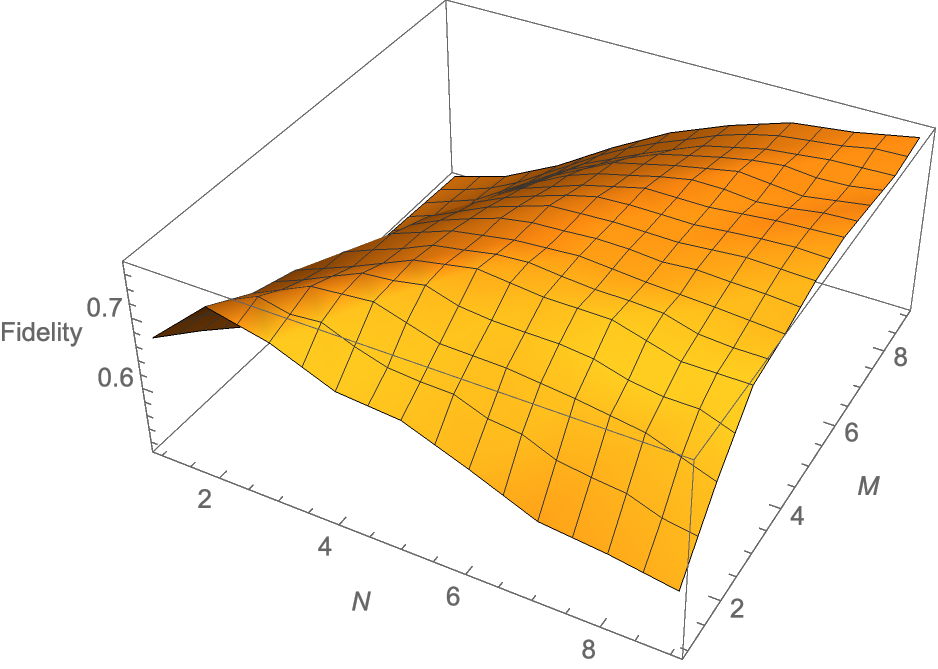}
\caption{\label{fig:Fidelity}Fidelity of counterfactual disembodied transport of an unknown qubit, counterportation, for a number of (CQZE) inner cycles N up to 10, and a number of (CQZE) outer cycles M up to 10, and realistic imperfections. These imperfections include Bob's trapped-atom qubit failing to reflect an impinging photon 34\% of the the time when it should reflect, and failing to not reflect 8\% of the time when it should not reflect. Counterportation fidelity for each choice of M and N is averaged over 100 evenly distributed qubits over the Bloch sphere, and violates the 2/3 classical fidelity limit even for the minimum number of outer- and inner-cycles, 2$\times$2.}
\end{figure}

\begin{figure}[ht]
\centering
\includegraphics[width=\linewidth]{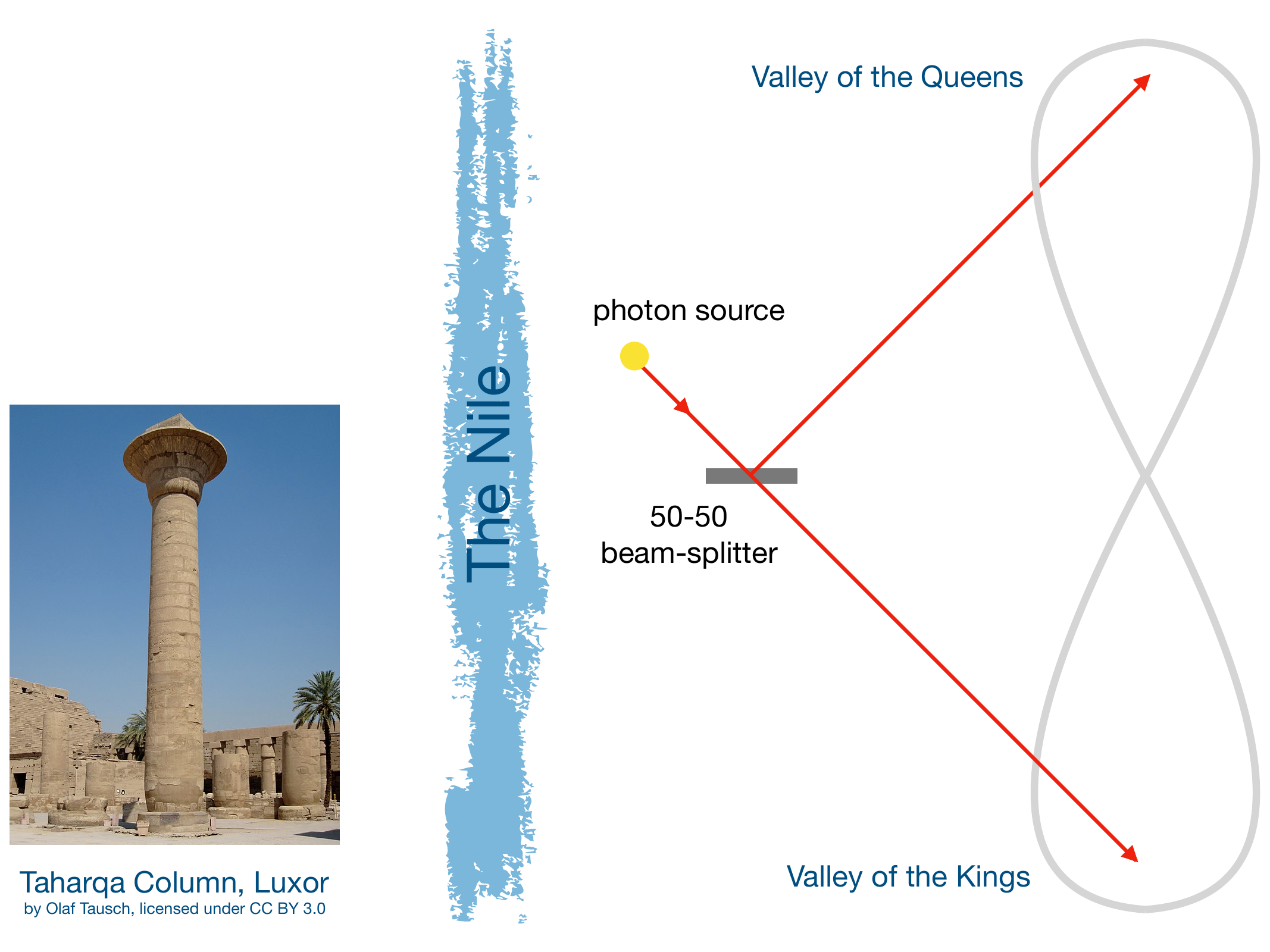}
\caption{\label{fig:SpatialEntanglement}Single-particle spatial entanglement. Consider a photon put in a superposition of taking two different paths near the bank of the River Nile, by means of a 50-50 beam-splitter. This is arranged such that at a later time the photon is in a superposition of being in the Valley of the Queens, and in the Valley of the Kings. We define a qubit for the photon being in the Valley of the Queens as $\left| \text{1} \right\rangle_Q$, and not being there as $\left| \text{0} \right\rangle_Q$. We similarly define another qubit for the photon being in the Valley of the Kings as $\left| \text{1} \right\rangle_K$, and not being there as $\left| \text{0} \right\rangle_K$. The overall state can thus be written, up to a normalisation factor, as $\left| \text{1} \right\rangle_Q\left| \text{0} \right\rangle_K+ \left| \text{0} \right\rangle_Q\left| \text{1} \right\rangle_K$, which is an entangled state across space. Counterportation relies on manipulating such spatial entanglements of a single particle into carefully choreographed self-interference scenarios---as in the scheme of Fig. \ref{fig:CQCNOT}---in order to render space disjunctly traversable: local wormhole.}
\end{figure}

Another approach to investigating where a photon has or has not been in the setup of Fig. \ref{fig:Paradox} is consistent histories\cite{Griffiths2016,Griffiths2002ConsistentTheory}, which we have employed in \cite{Salih2018a}. In the following we recreate the above discussed paradox and its resolution from the point of view of consistent histories. What is meant by a ``history'' here is a sequence of events between an initial state and a final state---a series of projections at various times during the system's unitary evolution. Each history has an associated chain-ket, whose inner product with itself assigns a probability to that particular history, making the consistent-histories approach time-symmetric. The idea is to construct a family of histories between the pre-selected state (in this case the photon in S, H-polarised at the start of an outer cycle) and the post-selected state (the photon in S, H-polarised at end of the outer cycle) that, first, has at least one history where the photon is in arm C, and second, the family is consistent, which means all histories are mutually orthogonal. For a consistent-history analysis it helps to think of measurements by detectors $D_A$ to take place after $t_{final}$, which is permitted by the deferred measurement principle\cite{Nielsen2010}. Here is the relevant family of consistent histories for the first outer cycle, 
\begin{align}
& S_0 \otimes H_0 \odot \left \{ A_1 \otimes I_1, D_1 \otimes I_1 \right \} \odot \left \{ A_2 \otimes I_2, B_2 \otimes I_2, C_2 \otimes I_2, \right \} \odot \nonumber \\ & \left \{ A_3 \otimes I_3, B_3 \otimes I_3, C_3 \otimes I_3, \right \} \odot S_4 \otimes H_4
\end{align}

where $S_0$ and $H_0$ are the projectors onto arm $S$ and polarisation $H$ at time $t_0$. $A_1$ and $I_1$ are the projectors onto arm $A$ and the identity polarisation $I$ at time $t_1$, and so on. The curly brackets contain different possible projectors at that particular time. There are 18 possible histories in this family. For example, the history ($S_0 \otimes H_0) \odot (A_1 \otimes I_1) \odot (A_2 \otimes I_2) \odot (A_3 \otimes I_3) \odot (S_4 \otimes H_4)$ has the photon travelling along arm A. Here's the chain ket associated with this history, $\left| S_0 \otimes H_0, A_1 \otimes I_1, A_2 \otimes I_2, A_3 \otimes I_3, S_4 \otimes H_4 \right\rangle = (S_4 \otimes H_4) T_{4,3} (A_3 \otimes I_3) T_{3,2} (A_2 \otimes I_2) T_{2,1} (A_1 \otimes I_1) T_{1,0} \left| \text{$S_0H_0$} \right\rangle$, where $T_{1,0}$ is the unitary transformation between times $t_0$ and $t_1$, $T_{2,1}$ is the unitary transformation between times $t_1$ and $t_2$, and so on. By applying these unitary transformations and projections, we see that this chain-ket is equal to, up to a normalisation factor, $\left| \text{$S_4H_4$} \right\rangle$. Other than the history with the photon in arm A, all other 17 histories have probability zero, including the ones where the photon is in arm C. For example the chain-ket $\left| S_0 \otimes H_0, D_1 \otimes I_1, C_2 \otimes I_2, C_3 \otimes I_3, S_4 \otimes H_4 \right\rangle = S_4 \otimes H_4 \left| \text{$J_4H_4$} \right\rangle$, up to a normalisation factor. Because projectors onto arms S and J are orthogonal, this chain-ket is zero. (Other chain-kets are zero because projectors onto polarisations H and V are orthogonal.) The photon was not in arm C during the first outer cycle, between times $t_0$ and $t_4$. 

Exactly the same goes for the second outer cycle, with the pre-selected state at time $t'_{0}$ and the post-selected state at time $t'_{4}$. Here is the relevant family of consistent histories for the second outer cycle,
\begin{align}
& S_{0'} \otimes H_{0'} \odot \left \{ A_{1'} \otimes I_{1'}, D_{1'} \otimes I_{1'} \right \} \odot \left \{ A_{2'} \otimes I_{2'}, B_{2'} \otimes I_{2'}, C_{2'} \otimes I_{2'}, \right \} \odot \nonumber \\ & \left \{ A_{3'} \otimes I_{3'}, B_{3'} \otimes I_{3'}, C_{3'} \otimes I_{3'}, \right \} \odot S_{4'} \otimes H_{4'}
\end{align}

Since all histories in this consistent family are also zero, except the one where the photon travels down arm A, the photon was not in arm C during the second outer cycle, between times $t'_{0}$ and $t'_{4}$. Therefore the photon was not in arm C at any time. %Rejecting this result, as did \cite{Griffiths2016}, is analogous to conceding that some statement about a physical system is true between times 1:00:00pm and 1:00:01pm, true, for the exact same physical system between 1:00:01pm and 1:00:02pm, but is somehow not applicable between 1:00:00pm and 1:00:02pm.

Interestingly, the paradox presented in the text can be reproduced using consistent histories. Given an initial pre-selected state with the photon at the source at the top of Fig. \ref{fig:Paradox}, H-polarised, and a final post-selected state of the photon in arm F on its way towards detector $D_0$ at the bottom, there exists a family of consistent histories that includes histories where the photon is in arm C during the second outer cycle, namely,
\begin{align}
& S_0 \otimes H_0 \odot \left \{ A_{1'} \otimes I_{1'}, D_{1'} \otimes I_{1'} \right \} \odot \left \{ A_{2'} \otimes I_{2'}, B_{2'} \otimes I_{2'}, C_{2 '}\otimes I_{2'} \right \} \odot \nonumber \\ & \left \{ A_{3'} \otimes I_{3'}, B_{3'} \otimes I_{3'}, C_{3'} \otimes I_{3'} \right \} \odot F_{final} \otimes H_{final} 
\end{align}

It is straightforward to check that this family is consistent, as each chain-ket is zero except the one associated with the history that has the photon in arm A. The photon was not in arm C during the second outer cycle. Now the analogous family that would allow us to ask of the whereabouts of the photon during the first outer cycle is,
\begin{align}
& S_0 \otimes H_0 \odot \left \{ A_{1} \otimes I_{1}, D_{1} \otimes I_{1} \right \} \odot \left \{ A_{2} \otimes I_{2}, B_{2} \otimes I_{2}, C_{2 }\otimes I_{2} \right \} \odot \nonumber \\ & \left \{ A_{3} \otimes I_{3}, B_{3} \otimes I_{3}, C_{3} \otimes I_{3} \right \} \odot F_{final} \otimes H_{final} 
\end{align}

This family however is not consistent, as its histories are not all mutually orthogonal. Besides the nonzero chain-ket associated with the history that has the photon in arm A, the chain-ket $\left| S_0 \otimes H_0, D_1 \otimes I_1, C_2 \otimes I_2, B_3 \otimes I_3, F_{final} \otimes H_{final} \right\rangle$ is also nonzero, rendering the question of whether the photon was in arm C during the first outer cycle meaningless within this framework. We therefore seem to have one conclusion based on consistent histories for the first outer cycle, but a different one for the exactly identical---as far as standard quantum mechanics is concerned---second outer cycle, which once more is unsettlingly paradoxical.

The paradox is resolved by considering each outer cycle {\it separately}, that is with the pre-selected state at the beginning of the outer cycle and the post-selected state at the end of the outer cycle, as shown above. Contrary to the objection in \cite{Griffiths2018}, this does not violate the single framework rule, which states that different consistent histories families (or frameworks) cannot be applied during the same time interval \cite{Griffiths2002ConsistentTheory}, which is categorically not the case here. Consider the intuitive approach given in Ch. 16 of reference \cite{Griffiths2002ConsistentTheory} for combining conclusions drawn based on two, even incompatible frameworks, ``The conceptual difficulty goes away if one supposes that the two incompatible frameworks are being used to describe... the same system during two different runs of an experiment.'' Since in the setup we are analysing, each outer cycle is identical, we are effectively looking at the same system during different runs of the experiment, using the same, therefore compatible, framework. The paradox is thus resolved. The photon was not in arm C during the first outer cycle. It was not in C during the second outer cycle. Therefore it was never in C.

Even for the inconsistent family arising from the pre-selected state being at the beginning of the first outer cycle and the post-selected state being at the end of the final outer cycle, the probability associated with any history that includes a projection on C is upper-bounded by $\sin^2 \frac{\pi}{2M}$, since it is an inner-cycle projection. M is the number of outer-cycles. Consequently, for $M>>1$ approximate consistency is satisfied as, intuitively, there can be no interference with such zero-probability histories \cite{Halliwell1993}. Once more, the photon was never in C.

\bibliographystyle{naturemag}
\bibliography{references.bib}

\begin{figure}[ht]
\centering
\includegraphics[width=\linewidth]{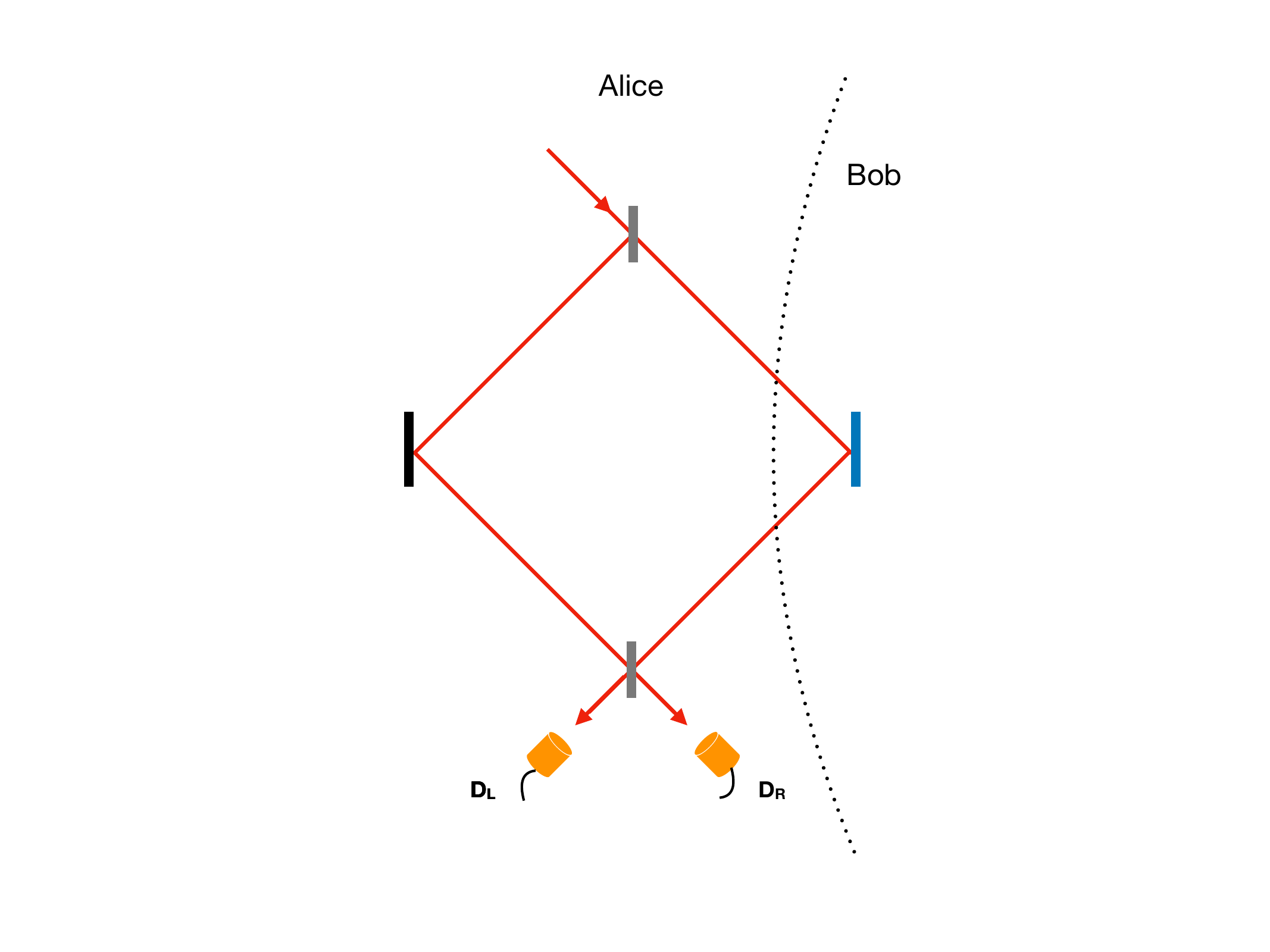}
\caption{\label{fig:SimplestCounterfactual}Bonus material: one of the simplest exchange-free protocols, if not the simplest. (Building on Figure 5 in arXiv:1807.06586v13, the protocol is a stripped-down version of \protect\cite{Salih2018b}.) The two 50-50 beam-splitters mean that a photon entering from the top always ends up at detector $D_R$, provided Bob's switchable mirror is turned on, causing it to reflect. Bob then encodes bit 0 by turning his switchable mirror on, and encodes bit 1 by turning it off. Alice starts by sending in a photon with probability p, where p is close to but not equal to 1. If she doesn't send a photon, probability 1-p, she records bit 0. If on the other hand she sends a photon, there are two scenarios. First, if Bob doesn't block his interferomer arm by turning on his mirror, then the photon goes straight to detector $D_R$ and the protocol fails. Second, if Bob blocks the channel by turning off his mirror, Alice's photon now has a finite chance (1/4) of triggering detector $D_L$, in which case Alice correctly records bit 1. If detector $D_R$ clicks the protocol fails. Bits corresponding to failed runs are discarded. Post-selecting on Alice recording a bit, the error probability when Bob doesn't block is 0, while the error probability when Bob blocks is (4-4p)/(4-3p). Therefore, as p approaches 1 the error probability approaches 0, but the probability of a successful run when Bob does not block also approaches 0. Importantly, given enough runs, there's a finite probability of an arbitrarily long message being sent with arbitrarily high accuracy.}
\end{figure}

%Suggested referees to include:
%1. Xiongfeng Ma, Tsinghua University, xma@tsinghua.edu.cn
%2. Onur Hosten, IST Austria, onur.hosten@ist.ac.at
%3. Yuan Cao, Tsinghua University, yuancao@ustc.edu.cn
%4. Arun Kumar Pati, Harish-Chandra Research Institute, akpati@hri.res.in
%5. Roger Colbeck, University of York, roger.colbeck@york.ac.uk

%Suggested referees to exclude from reviewing because of direct competition:

%1. Lev Vaidman, Tel Aviv University, vaidman@post.tau.ac.il
%2. Robert Griffiths, Carnegie Mellon University, rgrif@andrew.cmu.edu
%3. Suhail Zubairy, TAMU, zubairy@physics.tamu.edu
%4. Zheng-Hong Li, TAMU, refirefox@shu.edu.cn
%5. David Arvidsson-Shukur, MIT, drma2@cam.ac.uk

%A new physics paradox is brought to light. While interesting in its own right, its resolution underscores the possibility under the laws of Nature of what we coin counterportaion. A qubit can be transported across space without pre-shared entanglement, classical communication, or indeed any particles exchanging hands---using currently available, far-from-perfect, devices. This challenges some of the basic assumptions of teleportation, one of the cornerstones of quantum information. Remarkably, the phenomenon of counterportation provides a smoking gun for the existence of an underlying physical reality.

\end{document}